\documentclass[twocolumn]{aastex63}
\usepackage{xcolor}
\usepackage{amsmath}

\newcommand{\msun}{M_{\odot}}

\graphicspath{{./}{figures/}}

\begin{document}

\title{Fast Parameter Estimation of Binary Mergers for Multimessenger Followup}

\correspondingauthor{Daniel Finstad}
\email{dfinstad@syr.edu}

\author[0000-0002-8133-3557]{Daniel Finstad}
\affiliation{Department of Physics, Syracuse University, Syracuse, NY 13244, USA}
\author[0000-0002-9180-5765]{Duncan A. Brown}
\affiliation{Department of Physics, Syracuse University, Syracuse, NY 13244, USA}

\begin{abstract}
Significant human and observational resources have been dedicated to electromagnetic followup of gravitational-wave events detected by Advanced LIGO and Virgo. As the sensitivity of LIGO and Virgo improves, the rate of sources detected will increase. \cite{Margalit:2019dpi} have suggested that it may be necessary to prioritize observations of future events. Optimal prioritization requires a rapid measurement of a gravitational-wave event's masses and spins, as these can determine the nature of any electromagnetic emission. We extend the relative binning method of \cite{Zackay:2018qdy} to a coherent detector-network statistic. We show that the method can be seeded from the output of a matched-filter search and used in a Bayesian parameter measurement framework to produce marginalized posterior probability densities for the source's parameters within 20 minutes of detection on 32 CPU cores. We demonstrate that this algorithm produces unbiased estimates of the parameters with the same accuracy as running parameter estimation using the standard gravitational-wave likelihood. We encourage the adoption of this method in future LIGO-Virgo observing runs to allow fast dissemination of the parameters of detected events so that the observing community can make best use of its resources.
\end{abstract}


\section{Introduction} \label{sec:intro}

The observation of the binary neutron star merger GW170817 in gravitational and electromagnetic waves \citep{TheLIGOScientific:2017qsa,GBM:2017lvd} has demonstrated the importance of multimessenger astronomy in answering fundamental questions in physics, astronomy, and cosmology; see e.g.~\cite{Monitor:2017mdv}, \cite{Lattimer:2019iye}, and \cite{Abbott:2017xzu}. With the observation of GW190814, gravitational-wave astronomy has begun to explore the properties of compact objects that are more massive than previously observed neutron stars and less massive than previously observed black holes~\cite{Abbott:2020khf}. Advanced LIGO and Virgo perform a search for compact-object binary mergers with several low-latency analyses based on matched filtering \citep{Messick:2016aqy,Nitz:2018rgo} and release alerts to the astronomical community to enable followup of detected events. As the sensitivity of the Advanced LIGO and Virgo detectors improves, the rate at which interesting events are detected will increase. \cite{Margalit:2019dpi} have suggested that it may become necessary to prioritize events for followup in future LIGO-Virgo observing runs. Optimal prioritization will require the knowledge of the source-frame component masses and spins of the binary, as these determine the type of electromagnetic counterpart that may be generated by the merger \citep{Foucart:2018rjc,Capano:2019eae}.

In this paper, we demonstrate that it is possible to perform full Bayesian parameter estimation on binary neutron star and neutron star--black holes signals within 20 minutes of the source's detection by  a matched-filter search (with an average time of 10.8 minutes) using  32 CPU cores (2.5~GHz Xeon\textsuperscript{\textregistered} Gold 6248). Our analysis produces marginalized posterior probability densities for the source's parameters (including source-frame masses, spins, sky location, and distance) that can be used to guide the prioritization of electromagnetic followup in future LIGO-Virgo observing runs. We achieve this by extending the relative binning method of \cite{Zackay:2018qdy} to a fully coherent statistic, seeding the relative binning algorithm from the output of a matched-filter search, and using the \texttt{dynesty} nested-sampling package \citep{Speagle_2020}. We have made our code available in the PyCBC Inference framework \citep{Biwer:2018osg}.

We validate our analysis on a population of simulated binary neutron star and neutron star--black hole signals in a LIGO-Virgo detector network. A matched-filter search is used to identify signals that have a false alarm rate better than one per month. We then use our algorithm to produce marginalized posterior probability densities for each qualifying signal. For the parameters of interest, we perform a percentile-percentile test and demonstrate that our method produces unbiased parameter estimates. Comparing our sky localization to that of the \texttt{Bayestar} algorithm \citep{Singer_2016}, we find that the 90\% credible localization area improves by an average of 14 deg$^2$. We find that our analysis can recover the source-frame chirp mass to an accuracy of $\sim 5 \times 10^{-2}\msun$ for binary neutron star signals and $\sim 10^{-1}\msun$ for neutron star--black hole signals. We demonstrate that the measurement of mass ratio and spin is consistent with that of parameter estimation using the full likelihood, although these quantities are measured less accurately than the chirp mass as they enter the gravitational waveform at higher order and suffer from a partial degeneracy \citep{Cutler:1994ys,Hannam_2013}. As an example use case, we demonstrate that our method recovers essentially the same posterior probabilities for the parameters of GW170817 as the full likelihood calculation. Our method obtains marginalized posteriors for GW170817 in 20 minutes, compared to over three hours using the standard likelihood calculation.

This Letter is organized as follows: In Section \ref{sec:search} we describe our simulated search. Section \ref{sec:pe} describes our parameter estimation analysis and our implementation of relative binning for a detector network. Section \ref{sec:results} present our results including analysis run times and parameter estimation accuracy. Finally, we contrast our results to current methods in Section \ref{sec:conclusion}.

\section{Simulated Search}\label{sec:search}
We simulate a three-detector network representing the LIGO Hanford, LIGO Livingston \citep{TheLIGOScientific:2016agk,Buikema:2020dlj}, and Virgo \citep{TheVirgo:2014hva} detectors. We generate two populations of simulated signals: 600 binary neutron star and 570 neutron star--black hole binaries. Each population is injected into a realization of 33 hours of simulated detector data, which is created by coloring Gaussian noise to the design power spectral density of each detector \citep{Aasi:2013wya}. The simulated binary neutron star signals have their chirp mass drawn uniformly from the interval $[0.5, 3]~\msun$ and mass ratio $q=m_1/m_2$ drawn uniformly from the interval $[1, 3]$, with constraints on the component masses so that $1 < m_{1,2}/\msun < 3$. The neutron star's spins are restricted to be aligned with the orbital angular momentum and have dimensionless magnitude drawn uniformly from the interval $[-0.05, 0.05]$. The simulated neutron star--black hole signals have their chirp mass drawn uniformly from the interval $[0.5, 7]~\msun$ and mass ratio drawn uniformly from the interval $[1, 10]$, with constraints on the component masses so that $1 < m_{1,2}/\msun < 10$. Both component spins are restricted to be aligned with the orbital angular momentum, with the black hole spin dimensionless magnitude drawn uniformly from the interval $[-0.998, 0.998]$ and the neutron star spin dimensionless magnitude drawn uniformly from the interval $[-0.05, 0.05]$. This population of sources is chosen to cover the region in which it is expected that there will be neutron star disruption and an electromagnetic counterpart \citep{Capano:2019eae}. Each set of simulated signals is uniformly distributed in sky location and follow a uniform-in-volume distance distribution with $d_{L}\in[10, 300]$~Mpc for binary neutron star signals and $d_{L}\in[10, 500]$~Mpc for neutron star--black hole signals. Binary neutron star signals are simulated using the TaylorF2 waveform approximant~\cite{Dhurandhar:1992mw,Droz:1999qx,Blanchet:1995ez,Faye:2012we}. The neutron star--black hole signals are simulated using the IMRPhenomD approximant \citep{Husa:2015iqa,Khan:2015jqa}. For both populations, we set the tidal deformability of the neutron stars $\Lambda$ to zero, as this does not have a significant effect on the parameters we are investigating in this paper \citep{Damour:2012yf}.

To simulate the output of the LIGO-Virgo searches, we run each set of simulated signals through the PyCBC search pipeline \citep{Usman:2015kfa} configured to operate in a similar way to the PyCBC Live low-latency search used in the recent Advanced LIGO--Virgo observing runs \citep{DalCanton:2020vpm}. This search uses matched filtering \citep{Allen_2012} with a template bank of gravitational waveforms designed to give at least a 97\% match, measured by noise-weighted overlap, to any potential signal in the relevant parameter space \citep{DalCanton:2017ala,LIGOScientific:2018mvr}. The bank is designed to catch potentially electromagnetically-bright signals, and contains 315,325 waveforms. Template waveforms have component masses spanning [1, 30] $\msun$ and dimensionless spin magnitudes in the range [-1, 1], with the spin restricted to the direction of the orbital angular momentum. Templates in the bank are generated using the TaylorF2 approximant for signals with total mass $M=m_{1}+m_{2}< 4\msun$ \citep{Faye:2012we}, and with a reduced-order model of the SEOBNRv4 approximant otherwise \citep{Bohe:2016gbl}. Candidate triggers are required to be matched by the same template in at least two detectors in the network and with consistent phase, amplitude, and time of arrival given the network orientation and relative sensitivities between detectors \citep{Nitz:2017svb}. The search pipeline provides best-fit template parameters for every trigger and measures the trigger's statistical significance. The significance of a trigger is determined by the time-slide method and the pipeline computes a false alarm rate for each trigger. We select the triggers that have a false alarm rate more significant than 1 per month as candidate events for parameter estimation followup. This threshold is similar to that used to release low-latency events as public alerts for electromagnetic followup in the third LIGO-Virgo observing run \citep{emfollowdoc}. Of the total injections made, 306 binary neutron star and 253 neutron star--black hole injections satisfied this threshold.

\section{Parameter estimation}\label{sec:pe}
% %%%%%%% EQUATIONS FOR PE %%%%%%%%%%
% % general multi-ifo likelihood
% \begin{equation}
%     \mathcal{L}(d|\theta)\propto\exp\left[-\frac{1}{2}\sum_{i\in\mathrm{dets}}\left<d_{i}-h_{i}(\theta)|d_{i}-h_{i}(\theta)\right>\right]
% \end{equation}

% % general single-ifo likelihood
% \begin{equation}
%     \mathcal{L}(d|\theta)\propto\exp\left[-\frac{1}{2}\left<d-h(\theta)|d-h(\theta)\right>\right]
% \end{equation}

% % noise-weighted inner product definition
% \begin{equation}
%     \left<a|b\right>=4\mathfrak{R}\int_{f_{\mathrm{min}}}^{f_{\mathrm{max}}}\frac{\tilde{a}^{*}(f)\tilde{b}(f)}{S(f)}df
% \end{equation}

% % general posterior
% \begin{equation}
%     p(\theta|d)\propto\mathcal{L}(d|\theta)p(\theta)
% \end{equation}

% % single param posterior
% \begin{equation}
%     p(x|d)=\int \mathcal{L}(d|\mu,x)p(\mu, x)d\mu
% \end{equation}
% %%%%%%%%%%%%%%%%%%%%%%%%%%%%%%%%%%%

We use \textit{PyCBC Inference} \citep{Biwer:2018osg} with the \texttt{dynesty} nested sampler \citep{Speagle_2020} to perform Bayesian parameter estimation on candidate events from the search pipeline. In general, under the assumption of Gaussian noise characterized by a power spectrum $S(f)$, the likelihood of obtaining detector data $d$ given the presence of a gravitational waveform $h(\theta)$ is
\begin{equation}
    \mathcal{L}(d|\theta)\propto\exp\left[-\frac{1}{2}\left<d-h(\theta)|d-h(\theta)\right>\right],
\end{equation}
where
\begin{equation}
    \left<a|b\right>=4\mathfrak{R}\int_{f_{\mathrm{min}}}^{f_{\mathrm{max}}}\frac{\tilde{a}^{*}(f)\tilde{b}(f)}{S(f)}df
\end{equation}
is the noise-weighted inner product \citep{Finn:1992xs,Chernoff:1993th}.
In evaluating this likelihood, we can obtain estimates of the gravitational-wave parameters $\theta$ through the posterior probability distribution
\begin{equation}
    p(\theta|d)\propto\mathcal{L}(d|\theta)p(\theta),
\end{equation}
where $p(\theta)$ is the assumed prior probability distribution of the parameters. To calculate the likelihood, we use the relative binning method of \cite{Zackay:2018qdy}, which uses a linear interpolation across frequency samples over which the accumulated phase difference $\delta\phi$ between a fiducial waveform and nearby waveforms is less than a tunable threshold. This effectively downsamples the number of frequency points used to compute the likelihood, thereby speeding up the parameter estimation.

The implementation of relative binning proposed by \cite{Zackay:2018qdy} did not incorporate a coherent network detection statistic. We extend their method to include the extrinsic parameters which are needed to measure the sky location of an event: right ascension $\alpha$, declination $\delta$, geocentric time of coalescence $t_c$, inclination angle $\iota$, and gravitational-wave polarization angle $\psi$. These parameters are incorporated into the likelihood by projecting each template waveform onto the individual detectors in the network. A general frequency domain waveform template $h$ as seen by a detector can be written as
\begin{equation}
    h(f)=F_{+}(\alpha,\delta,\psi)h_{+}(f) + F_{\times}(\alpha,\delta,\psi)h_{\times}(f)
\end{equation}
where $h_{+,\times}$ are the plus and cross polarizations of the waveform, and $F_{+,\times}$ are the detector antenna responses to the two polarizations \citep{Anderson_2001}. The amplitude of the individual waveform polarizations depend on the inclination angle $\iota$ \citep{thorne.k:1987}
\begin{equation}
    h_{+}\propto \frac{1}{2}(1+\cos^{2}{\iota}),
\end{equation}
\begin{equation}
    h_{\times}\propto \cos{\iota}.
\end{equation}
We generate waveforms using both polarizations in order to capture this dependence. Similarly, we measure $\alpha$, $\delta$, $t_{c}$, and $\psi$ dependence through the detector antenna responses as the orientation of the detector arms, and thus the sensitivity to the two polarizations, will change as the Earth moves. To account for coherent network timing delays, we calculate detector-specific arrival times for each template waveform using $\alpha$, $\delta$, and $t_{c}$, based on the geometry of the network with respect to the source at the time of the signal, along with the light travel time from the Earth center \citep{Fairhurst:2009tc}.

The relative-binned likelihood calculation requires a fiducial waveform known to be near the peak of the likelihood. The chirp mass of the template used to generate a candidate by a search pipeline is accurate to within a few $10^{-3} \msun$ for binary neutron star signals \citep{Biscoveanu_2019} and to approximately $1\%$ for neutron star--black hole signals \citep{canton2020realtime}. Since the chirp mass is the leading order parameter governing phase evolution for a binary inspiral \citep{Peters:1963ux}, the best-fit template will be near the peak of the likelihood. We therefore use the parameters that the search pipeline reports for a signal to generate the fiducial waveform that seeds the relative binning method. For the fiducial sky location, inclination, and polarization, we arbitrarily choose $\alpha_{f}=\pi$, $\delta_{f}=0$, $\iota_{f}=0$, and $\psi_{f}=\pi$, as we find that more accurate initial estimates are unnecessary to correctly recover the source parameters. The fiducial coalescence time is set to be the arithmetic mean of the coalescence time reported by the search pipeline for each detector.

Parameter estimation is performed over the detector-frame chirp mass $\mathcal{M}$, the mass ratio $q=m_{1}/m_{2},~m_{1}\ge m_{2}$, the component aligned spins $\chi_{1,2}$, the geocentric time of coalescence $t_{c}$, the inclination angle $\iota$, the right ascension $\alpha$, the declination $\delta$, the luminosity distance $d_{L}$, and the gravitational-wave polarization angle $\psi$. The likelihood calculation includes an analytic marginalization over the coalescence phase $\phi_{c}$. We use the TaylorF2 approximant to generate the likelihood for binary neutron star waveforms and the IMRPhenomD approximant for the neutron star--black hole waveforms.

The prior distributions used in the parameter estimation are the same as those of the corresponding population of simulated signals for each parameter, with the exception of the chirp mass which we restrict to be uniform in $\mathcal{M} \in [\mathcal{M}_{s}-0.1, \mathcal{M}_{s}+0.1]~\msun$, where $\mathcal{M}_{s}$ is the chirp mass of the template reported by the search. This constraint on the chirp mass prior enables quicker convergence of the parameter estimation, but in all cases the restricted bounds are well outside the region of posterior support and so do not affect the accuracy of recovery.

For each simulated signal recovered with false alarm rate more significant than 1 per month by the search pipeline, we run the relative-binned parameter estimation analysis
to produce posterior distributions for the 10-dimensional set of waveform parameters $\theta = (\mathcal{M}, q, \chi_{1}, \chi_{2}, t_{c}, \iota, \alpha, \delta, d_{L}, \psi)$. For each signal, we measure the wall-clock time that it takes to perform the parameter estimation on 32 cores of an Intel\textsuperscript{\textregistered} Xeon\textsuperscript{\textregistered} Gold 6248 CPU running at a clock speed of 2.5~GHz. 

\section{Results} \label{sec:results}

\begin{figure*}[ht]
\includegraphics[width=\textwidth]{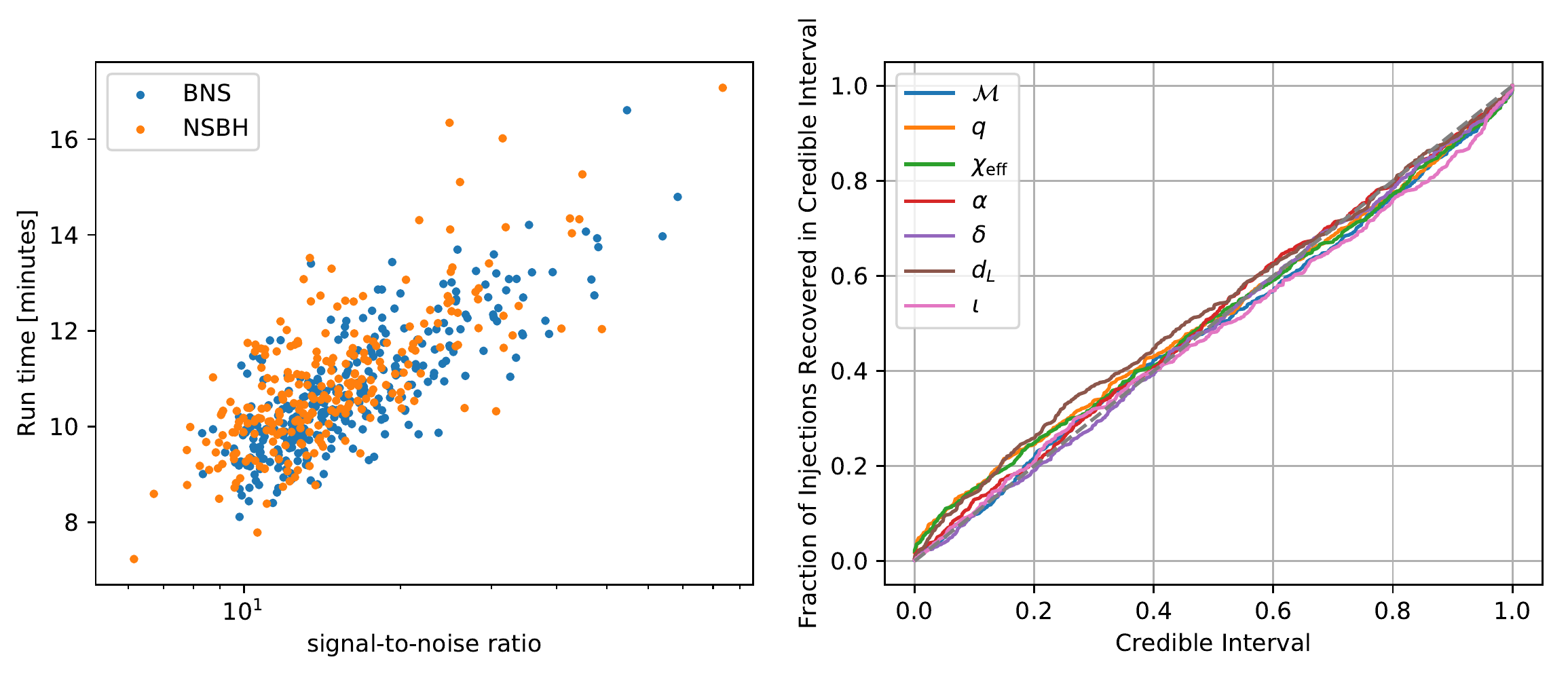}
\caption{\textit{Left:} The wall-clock time in minutes that it takes to perform the parameter estimation using the coherent relative binning likelihood and nested sampling on 32 cores of an Intel\textsuperscript{\textregistered} Xeon\textsuperscript{\textregistered} Gold 6248 CPU running at a clock speed of 2.5~GHz as a function of the network signal-to-noise ratio of the maximum likelihood template. The average run-time for a single signal is 10.8 minutes, with the maximum run-time being 20 minutes for all signals. Increasing the number of cores does not significantly decrease the wall-clock run-time. The run-time shows a slight increase as a function of the signal-to-noise ratio, as expected given that signals with a larger signal-to-noise ratio have a more narrowly peaked likelihood. \textit{Right:} The fraction of injections recovered within a credible interval plotted as a function of credible interval. Fidelity to the 1:1 diagonal line is an indication of probability being uniformly distributed across a given parameter's posterior distribution and is a measure of the accuracy of this analysis at the population level. We find that all of the parameters of interest are estimated in an unbiased way by our parameter estimation method.}
\label{fig:timing_and_pp}
\end{figure*}

The timing results for the two simulated populations as a function of the network signal-to-noise ratio of the maximum likelihood template are shown in the left panel of Fig.~\ref{fig:timing_and_pp}. The average run time for a single signal is 10.8 minutes, with the maximum run-time being 20 minutes for all signals.  The parallelization used by the nested sampling algorithm is saturated at approximately 32 cores, so while a small decrease in wall-clock time may be gained by fine-tuning the number cores, adding additional cores beyond 32 does not significantly decrease the run time. Processor cores with a faster clock speed will generally decrease run time, however.

To determine whether our method of measuring the parameters is accurate for the population of injected signals, we perform a percentile-percentile (PP) test on each of the main parameters of interest: chirp mass $\mathcal{M}$, mass ratio $q$, effective spin $\chi_{\mathrm{eff}}=(m_{1}\chi_{1}+m_{2}\chi_{2})/(m_{1}+m_{2}$, right ascension $\alpha$, declination $\delta$, luminosity distance $d_L$, and inclination $\iota$. The PP test calculates the distribution of percentile ranks for all injected parameter values within their respective posteriors and constructs the fraction of injections recovered within a credible interval as a function of credible interval. Any deviation from uniformity in this distribution for a parameter is an indication of measurement bias. We measure any deviation with the Kolmogorov-Smirnov (KS) test \citep{10.2307/2280095}, which computes the distance between the empirical distribution that we find for the PP test and the expected distribution. The results of the PP tests are shown in the right panel of Fig.~\ref{fig:timing_and_pp}. For every parameter of interest, we find that the PP test follows the ideal distribution well, with the KS test indicating that the percentile rank distributions cannot be meaningfully distinguished from uniform. Our results show that our analysis produces unbiased estimates for each of the parameters of interest.

\begin{figure}[ht]
\includegraphics[width=0.45\textwidth]{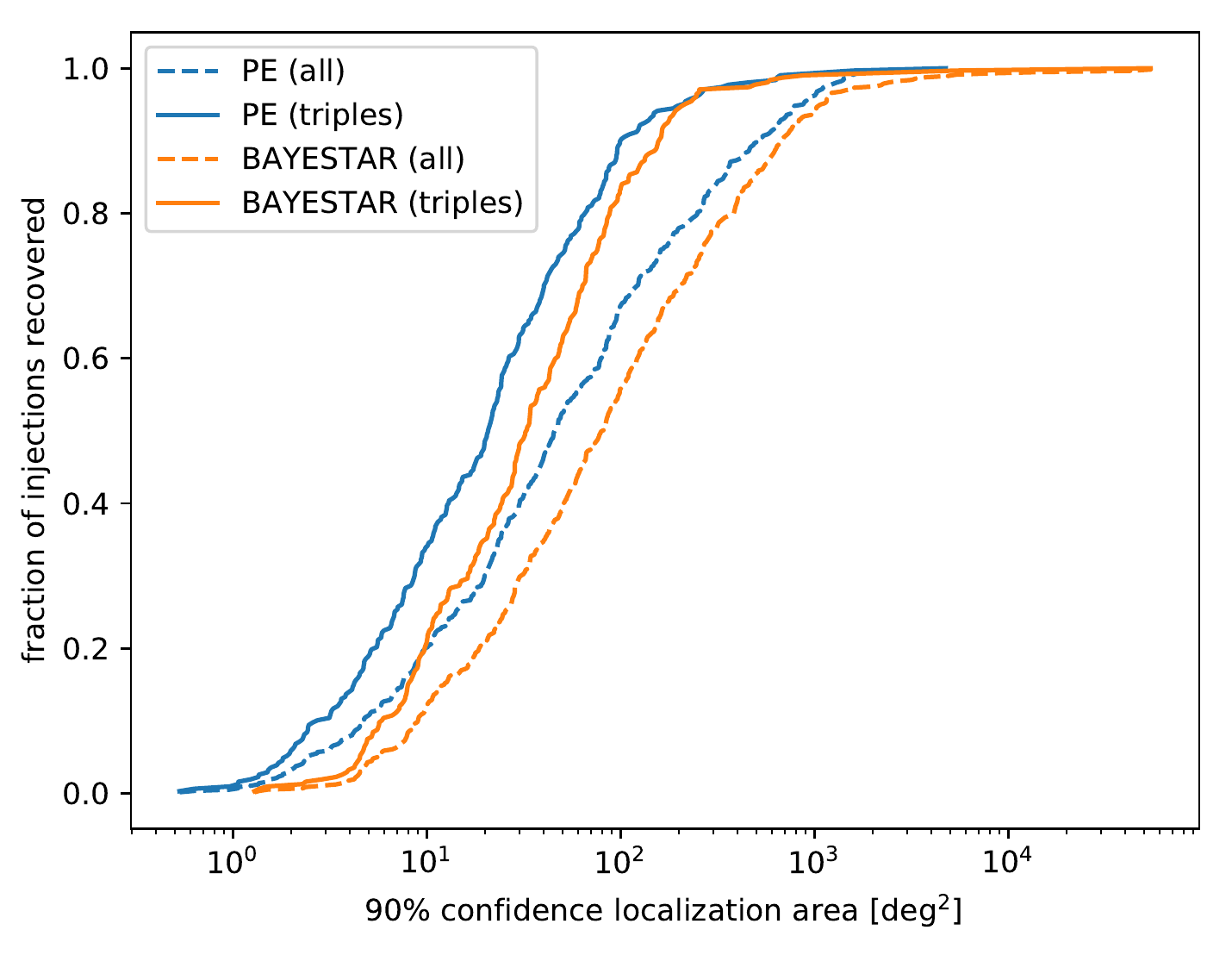}
\caption{Fraction of injections recovered as a function of the 90\% confidence localization area. The localization results from our parameter estimation analysis are shown in blue, and those from the \texttt{Bayestar} algorithm are in orange. The dotted lines show the results for the entire set of signals, while solid lines show only signals that were above the detection threshold in all three detectors in our simulated search. We find our localization areas are consistently smaller than those from \texttt{Bayestar}, as indicated by the blue lines lying to the left of the orange lines, although the difference in areas is not large. The improvement in localization area between \texttt{Bayestar} and our analysis is 14 deg$^2$ on average, and is comparable between the set of triple-coincident signals and the set of all signals.}
\label{fig:localization}
\end{figure}

To examine the accuracy of sky localization, we calculate the area on the sky containing 90\% of the probability for the location of the source. We compare the area of this probability contour to the 90\% credible interval of the sky-map produced in low-latency by the \texttt{Bayestar} algorithm \citep{Singer_2016}. Fig.~\ref{fig:localization} shows the cumulative fraction of signals recovered as a function of the 90\% confidence localization area for our method and by \texttt{Bayestar}. For direct comparison to the results of \cite{Singer_2016}, we calculate the cumulative fraction using all recovered signals, and the subset of the recovered signals that is detected above threshold in all three detectors.  We find that the area of the 90\% credible region improves by an average of 14 deg$^2$ when using the relative binning parameter estimation compared to \texttt{Bayestar}.

\begin{figure*}[ht]
\includegraphics[width=\textwidth]{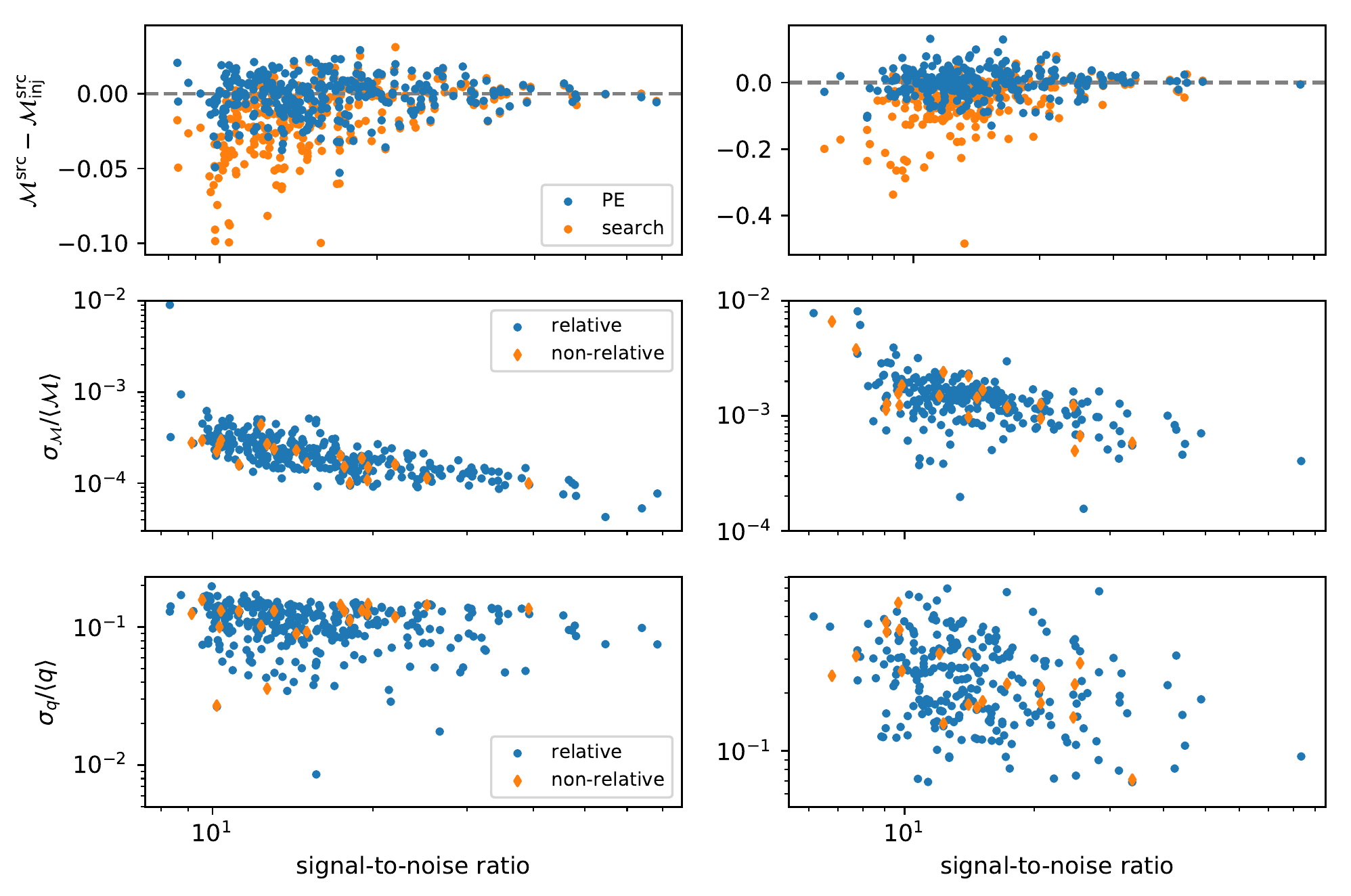}
\caption{Chirp mass and mass ratio recovery metrics for the binary neutron star (left column) and neutron star--black hole (right column) signals in our analysis. \textit{Top row:} Difference between source-frame chirp mass estimates and the true injected value, as a function of signal-to-noise ratio. Blue circles denote differences from the median posterior values from parameter estimation, while orange circles show differences from best-fit template values from the search. We find that on average our parameter estimation results improve on the accuracy of the best-fit template by a factor of $2$. \textit{Middle and bottom rows:} Fractional uncertainties on chirp mass and mass ratio, respectively, calculated as the ratio of standard deviation and mean of the posterior distributions. Uncertainties from our relative-binned analysis are shown as blue circles, and those from a standard non-relative likelihood analysis on a subset of the population are shown as orange diamonds. Our relative-binned results are consistent with the non-relative analysis, and also with the results in \cite{Farr:2015lna}.}
\label{fig:residuals_and_uncertainty}
\end{figure*}

To examine the accuracy of parameter recovery, we calculate the difference between the median of the posterior and the known injected value for each parameter. The accuracy of chirp mass recovery in the source-frame is shown in the top panels of Fig.~\ref{fig:residuals_and_uncertainty} as a function of the network signal-to-noise ratio for each recovered signal. As expected, the accuracy of recovery increases as the signal-to-noise increases. For binary neutron star signals the difference between the median value of the chirp mass posterior and the injected value is less than $\sim 5 \times 10^{-2}\msun$ for all simulated signals. This accuracy improves by a factor of $2$ for signal-to-noise greater than $20$. Neutron star--black hole signals generally have larger uncertainties on their parameters and we find chirp mass residuals for these signals to be less than $10^{-1}\msun$ for signal-to-noise greater than $10$ and a factor of $2$ less than that for signal-to-noise greater than $20$. For comparison, we show the accuracy of the source-frame chirp mass of the best-fit template from the search. The search measures the detector-frame parameters of the gravitational-wave signal, so we convert this to the source-frame by computing the redshift at the median distance reported by \texttt{Bayestar} for the candidate event. The accuracy of the best-fit chirp mass from the search is an order of magnitude worse than estimated by \cite{Biscoveanu_2019}. However, the majority of the error comes from the calculation of the source-frame chirp mass. Comparing the detector-frame chirp mass of the simulated signal to the best-fit template, we find errors of $\sim 10^{-3}\msun$. 

The middle row of Fig.~\ref{fig:residuals_and_uncertainty} shows the fractional uncertainty in the chirp mass $\sigma_\mathcal{M} / \langle \mathcal{M} \rangle$, where $\sigma_\mathcal{M}$ and $\langle \mathcal{M} \rangle$ are the standard deviation and mean of the posterior distribution. By this measure we find the accuracy of our method for the binary neutron star population is comparable to that of \cite{Farr:2015lna}. As an additional check we also run parameter estimation using the full likelihood for a subset of the population and find that the accuracy of the relative binning method is consistent with results using the full likelihood for both binary neutron star and neutron star--black hole signals. These results show that our recovery of the chirp mass for all signals has more than sufficient accuracy to determine the expected type of electromagnetic counterpart and the possible fate of the merger remnant using the method of \cite{Margalit:2019dpi}.

The gravitational-wave phase evolution is less sensitive to changes in the mass ratio  and so the component masses of the binary are less well recovered than the chirp mass \citep{Cutler:1994ys}. A degeneracy exists between the mass ratio and component spins of the binary which makes measuring the component masses and spins challenging, especially for neutron star--black hole systems \citep{Hannam_2013}. The bottom row of Fig.~\ref{fig:residuals_and_uncertainty} shows the accuracy of measuring the mass ratio $q = m_1/m_2$. Although the measurement of this parameter is less accurate than that of the chirp mass, our results are again comparable to those seen by \cite{Farr:2015lna} and they are consistent with our comparison analysis using the full likelihood on a subset of the population. This demonstrates that the reduced accuracy is intrinsic to the measurability of the parameter and not a result of using the relative binning algorithm.

\begin{figure*}[ht]
\includegraphics[width=\textwidth]{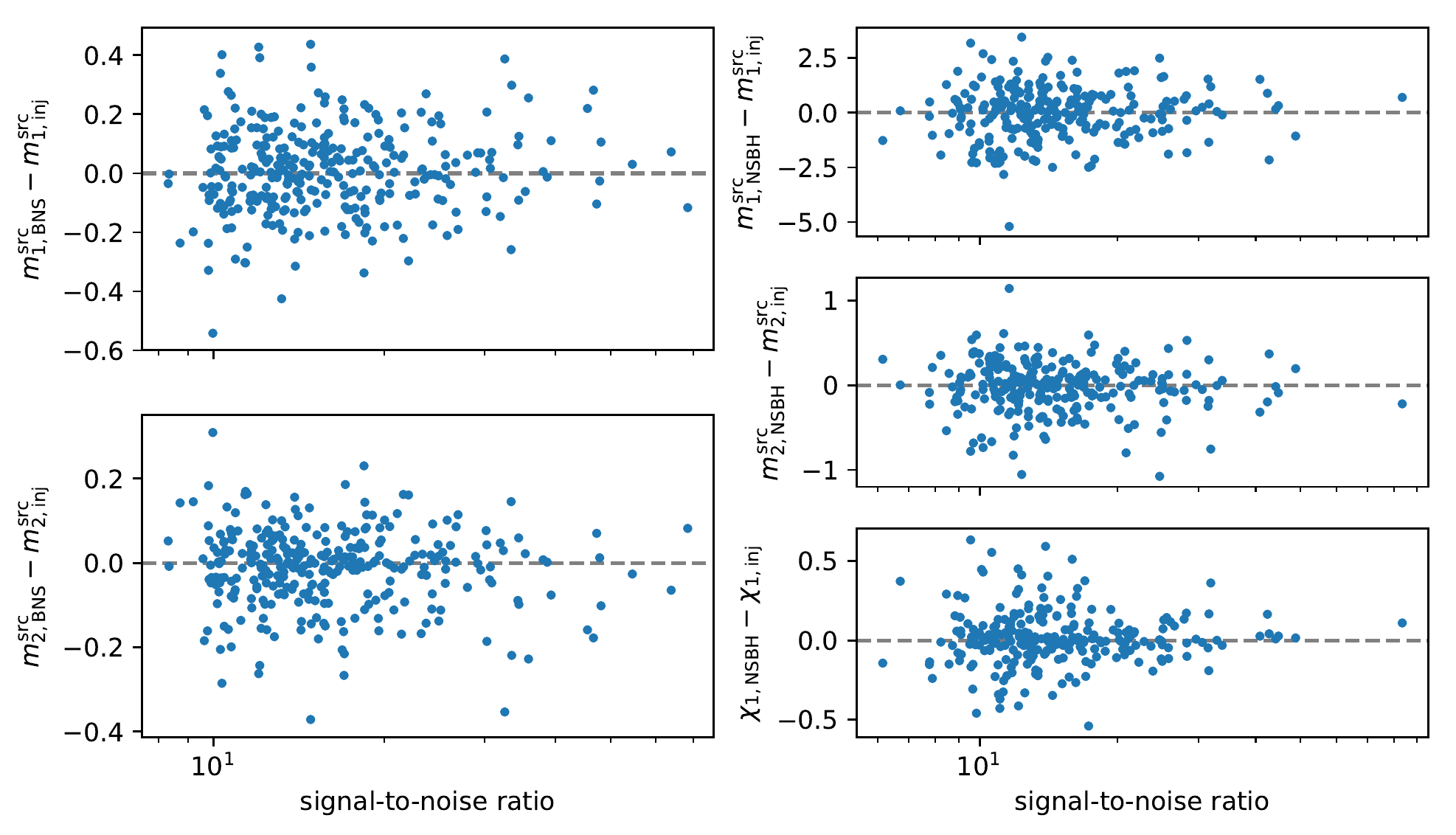}
\caption{Difference between parameter estimates and true injected values for some component parameters of interest, plotted against signal-to-noise ratio. The left column shows results for the component masses of binary neutron star signals, and the right column shows results for the component masses and black hole spin of neutron star--black hole signals. Differences are computed from median posterior values, and masses have been converted to the source-frame using the distance posteriors. We find both component masses of binary neutron star signals are generally constrained to within $\sim 0.5\msun$ of the true value for all signals, while the majority of primary and secondary masses of neutron star--black hole signals are within about $3\msun$ and $1\msun$ respectively. We find our black hole spin measurements are uninformative below a signal-to-noise of $20$, but for louder signals the spin is within about $0.3$ of the true value.}
\label{fig:component_recovery}
\end{figure*}

To further illustrate the utility of our method in recovering parameters of interest to the observing community, Fig.~\ref{fig:component_recovery} shows the source-frame component mass residuals for all signals as well as the black hole spin residuals for the neutron star--black hole signals, plotted as a function of the network signal-to-noise ratio. The binary neutron star component masses are shown in the left panels of the figure. The residuals on the primary mass are generally less than about $0.5\msun$ with only a slight tendency to smaller values as signal-to-noise increases. The secondary mass residuals are somewhat smaller, less than about $0.3\msun$, which can be attributed to the relatively narrow mass parameter space ($1-3\msun$) and our convention requiring $m_{2}<m_{1}$.

Neutron star--black hole signals have larger uncertainties on their intrinsic parameter estimates owing to the larger mass and spin parameter space and the known degeneracy between mass ratio and spin \citep{Hannam_2013}. However, these quantities are important in determining whether a merger will produce an electromagnetic counterpart. The residuals on component masses and black hole spin for our neutron star--black hole signals are shown in the right panels of Fig.~\ref{fig:component_recovery}. We find the primary and secondary mass residuals are mostly less than $3\msun$ and $1\msun$, respectively. Our estimates of the black hole spin are generally uninformative below a signal-to-noise of $20$, but above this threshold we find the residuals are constrained to be less than $\sim 0.3$.

\begin{figure*}[ht]
\includegraphics[width=\textwidth]{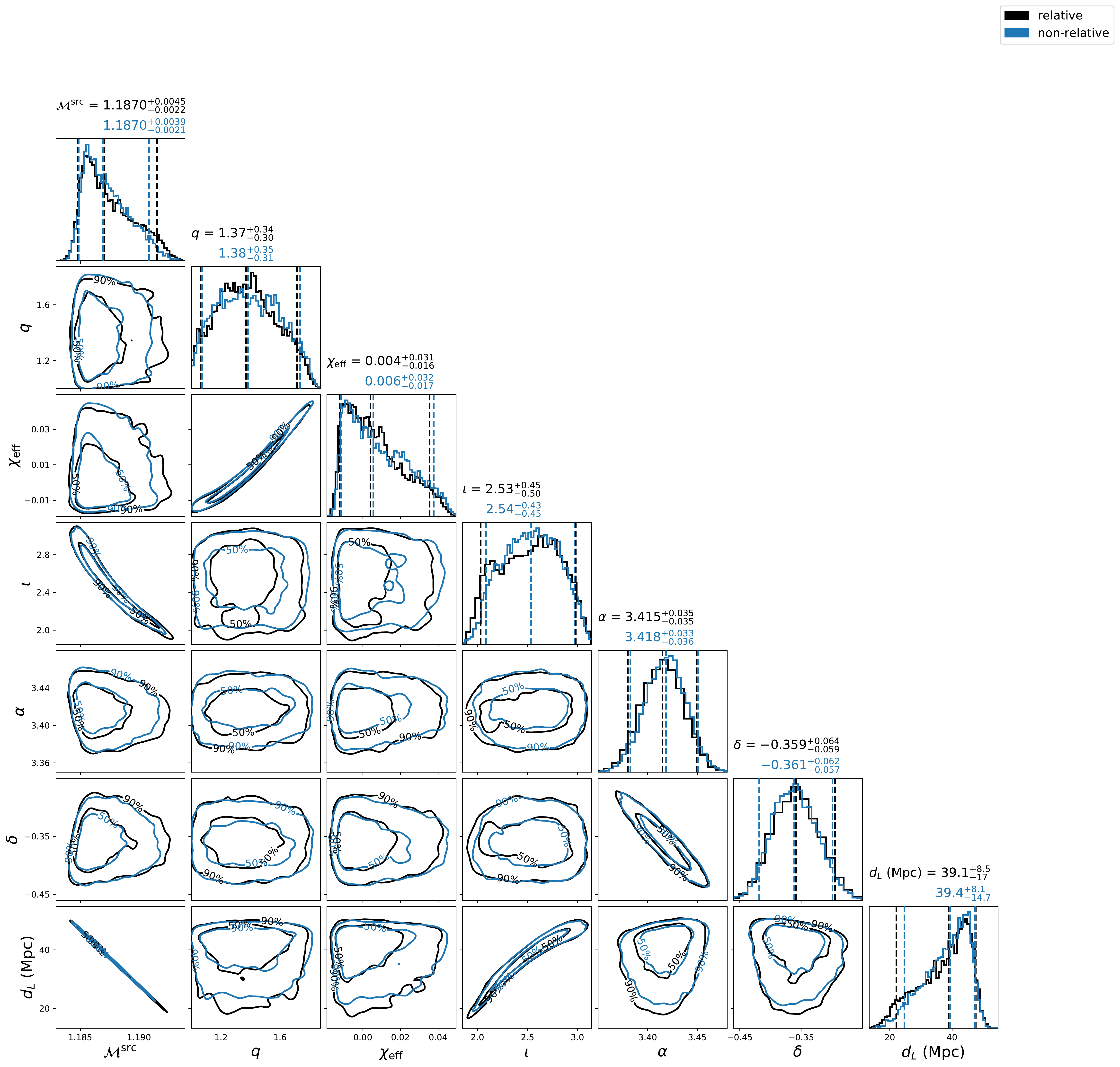}
\caption{Posterior distributions from a relative-binned parameter estimation analysis of GW170817 (black contour) as compared to a run using the standard non-relative likelihood (blue contour). Marginalized 1-dimensional histograms for each parameter are shown along the diagonal, with vertical dashed lines at the median value and the bounds of the 90\% credible interval. Off-diagonal plots show 2-dimensional slices of the parameter space with contours delineating the 50\% and 90\% credible regions. The relative-binned analysis completed in 20 minutes versus roughly 3 hours in the non-relative case, and all parameter distributions are consistent between the two analyses.}
\label{fig:gw170817_comparison}
\end{figure*}

As a final example of the effectiveness of our method, we apply it to GW170817 \cite{TheLIGOScientific:2017qsa} without including any prior knowledge of host galaxy location or distance. For comparison, we also repeat the analysis using a standard non-relative likelihood, and the posteriors from both runs are shown in Fig.~\ref{fig:gw170817_comparison}. For all measured parameters, we find the posterior distributions from the relative and non-relative analyses are nearly identical, in agreement with \cite{Dai:2018dca}. However, the analysis using the relative binning likelihood seeded by a search took only 20 minutes to complete, as compared to over 3 hours for the standard likelihood computation. In the only confirmed observation of a multimessenger gravitational-wave source to date, our analysis is able to provide the same localization region as the standard likelihood as well as the same intrinsic parameter estimates in substantially less computational time.

\section{Conclusion}\label{sec:conclusion}

In previous LIGO-Virgo observing runs, the information provided in low-latency to astronomers consisted of the time of the signal, an estimate of its statistical significance (false alarm rate), and a three-dimensional localization probability in sky location and distance. In the recent third observing run, two additional classifications were released that bin events into one of five broad categories (binary neutron star, binary black hole, neutron star--black hole, mass gap, or terrestrial noise) and estimate the probability that the event produced an electromagnetic counterpart \citep{Kapadia:2019uut,canton2020realtime}. Both of these methods are based on the parameters of the best-fit matched filter template recorded by the low-latency search. \cite{Biscoveanu_2019} performed a template-bank simulation that estimated that the low-latency chirp mass point estimate for binary neutron stars is accurate to $\sim 10^{-3}~\msun$, however they note that there can be significant bias in mass ratio and effective spin from the best-fit template. \cite{canton2020realtime} demonstrated that the best-fit chirp mass from a search can be used to inform a classification scheme in which the classifications are correct in a large majority of cases.

Here, we have extended the relative binning algorithm \citep{Zackay:2018qdy} for fast likelihood evaluation in gravitational-wave parameter estimation to a fully coherent detector network and demonstrated that it can be seeded by the output of a matched-filter search. We have applied our method to a set of 559 simulated signals (306 binary neutron star and 253 neutron star--black hole binaries) as well as to GW170817. We find that in all cases our method produces unbiased estimates for all measured parameters in less than 20 minutes. We have shown that our method is capable of producing full posterior distributions for all signal parameters, which do not suffer from the biases seen when attempting to measure the mass ratio and spin from the best-fit template. In the  case of GW170817, the relative-binned analysis produces results nearly identical to those from a standard analysis using the full likelihood, emphasizing our method's utility in producing fast parameter estimates that are of particular interest for electromagnetic followup.

For gravitational-wave events in LIGO's third observing run, the average time between an initial trigger alert and the first Bayesian parameter estimation results being made available was about 10 hours (although only updated sky maps are released and not measurement of the source's parameters). We have demonstrated  our method could reduce this delay time considerably, which would allow for electromagnetic followup campaigns to be conducted more efficiently.   We encourage the LIGO Scientific and Virgo collaborations to adopt these methods to provide the observing community with fast and accurate estimates of the parameters of detected signals so that these can be used to inform and prioritize electromagnetic followup strategies. Finally, we note that given the computational cost, very few large scale injection studies of low-mass gravitational-wave signals have been done. Our implementation of the relative binning method into \textit{PyCBC Inference} brings these sorts of studies within reach for even modestly equipped computing facilities.

\acknowledgments

We thank Brian Metzger, Ben Margalit, and Edo Berger for helpful discussions. The authors are supported by National Science Foundation Grant No. PHY-1707954 and PHY-2011655. DAB thanks the Kavli Institute for Theoretical Physics for hospitality and partial support from National Science Foundation Grant No. PHY-1748958. DF acknowledges support from a Syracuse University Research Excellence Doctoral Fellowship. Computational work was supported by Syracuse University and National Science Foundation award OAC-1541396. Supporting data for this manuscript is available from \url{https://github.com/sugwg/rapid-relbin-pe}.

%\appendix
%\section{Appendix information}

%\bibliography{references}{}
%\bibliographystyle{aasjournal}

\end{document}